\documentclass{PoS}

\title{Testing General Relativity}

\ShortTitle{Testing General Relativity}

\author{Angelo Tartaglia\thanks{Member of the GINGER collaboration.}\\
        Politecnico di Torino and INFN, Torino, Italy\\
        E-mail: \email{angelo.tartaglia@polito.it}}

\abstract{This lecture will present a review of the past and present tests of the General Relativity theory.\
          The essentials of the theory will be recalled and the measurable effects will be listed and analyzed.\
          The main historical confirmations of General Relativity will be described. Then, the present situation will be reviewed presenting a number of examples. The opportunities given by astrophysical and astrometric observations will be shortly discussed. Coming to terrestrial experiments the attention will be specially focused on ringlasers and a dedicated experiment for the Gran Sasso Laboratories, named by the acronym GINGER, will be presented. Mention will also be made of alternatives to the use of light, such as particle beams and superfluid rings.}

\FullConference{Gran Sasso Summer Institute 2014 Hands-On Experimental Underground Physics at LNGS - GSSI14,\\
		22 September - 03 October 2014\\
		INFN - Laboratori Nazionali del Gran Sasso, Assergi, Italy }

\begin{document}

\section{Introduction}

The General Relativity theory (GR) is a powerful intellectual endeavour whose conceptual structure was elaborated by Einstein in a ten years long effort, until he could present his field equations for gravity at the Prussian Academy of Science in November 1915 \cite{einstein1}, then publish the foundation of the general theory in 1916 \cite{einstein2}.

At the beginning there was very little (practically no) experimental need for a new theory of gravity: just a tiny residue in the accounting for the precession of the perihelion of Mercury. GR, however, allowed for a number of new interesting predictions part of which were soon verified as corresponding to actual phenomena, starting the long season of the tests of General Relativity. For long the most spectacular successes of GR were harvested in cosmology, giving origin to a whole new branch of physics: the relativistic cosmology.

In the following I shall just mention the historical tests of GR and shall not enter the domain of relativistic cosmology, even though it is from there that major demands for a revision or extension of GR come since almost a couple of decades. We shall rather focus on present, actual or possible, tests of the theory, become possible thanks to recent spectacular improvements of various technologies. Not excluding observations inside our galaxy, the main attention will be paid to experiments performed or feasible in the terrestrial environment and in particular in terrestrial laboratories.

The main difficulty with the verification of General Relativity is that usually the sought for effects are extremely weak, thus challenging all available technologies. Despite this elusiveness the interest in GR experimental tests is strong and even growing because of the mentioned feeling that something may be missing from the theory or at least require completion. GR is also puzzling at the other end of the scale of energies, where gravity becomes as strong as the other fundamental forces of nature; there we fall back into cosmology and meet the other big paradigm of modern physics: Quantum Mechanics. The most difficult and profound challenge is indeed the persisting incompatibility between Quantum Mechanics and General Relativity, which emerges at the Planck scale. This problem is however out of the scope of the present lecture. Here I shall review and expound what the state of the art on GR tests is and what the perspectives for the near future are, including an experiment whose feasibility in the Gran Sasso National Laboratories is presently under practical scrutiny.

\section{General Relativity at a glance}
GR is a non-trivial theory requiring refined mathematical tools, such as tensors, and deserving careful study. Those who are not acquainted with the basics of the theory should resort to fundamental books such as \cite{MTW} or other equally good texts. Here I will only review the essentials of GR, assuming all technical details as known, and focusing on what is needed for understanding the ratio of the experiments.

In the GR paradigm, nature consists of two separate ingredients: matter/energy and space-time. GR focuses on space-time and its properties, assuming it is a Riemannian four-dimensional manifold with Lorentzian signature. Matter/energy induces curvature in the manifold and curvature is what we usually call 'gravitational field'. Gravity is so a geometrical property of space-time and GR is essentially geometry. Special Relativity (SR) is incorporated into GR via the Lorentz signature that insures the local speed of light to be the same $c$ for all freely falling observers; geometrically SR is a local approximation of GR, in the sense that Minkowski space-time is the local \textit{tangent space} of the full curved Riemannian manifold.

All relevant physical quantities are expressed by tensors, which are the right mathematical tools in order to formulate the physical laws in \textit{covariant} form, i.e. in a way independent from the arbitrary choice of the reference frame and the coordinate system. The seminal geometric object of GR is the (squared) \textit{line element} $ds^2$ representing the elementary 'distance' (interval) between two arbitrarily near positions (events) in space-time. After choosing a coordinate system it is\footnote{Einstein summation convention is assumed. Greek indices run from 0 to 3; Latin indices run from 1 to 3.}:
\begin{equation}
ds^2=g_{\mu\nu}dx^{\mu}dx^{\nu}
\end{equation}
The $g_{\mu\nu}$'s are the elements of the (symmetric) \textit{metric tensor} which incorporates the geometric properties of the manifold. Starting from the metric tensor it is possible to build a number of important geometric objects. From the $g_{\mu\nu}$'s and their first order derivatives one can generate the Christoffel symbols (the \textit{connection}), which do not correspond to a tensor, but are essential for building the \textit{covariant} derivatives, i.e. for accounting for the global change of any physical quantity when moving from one place to another in space-time: both the intrinsic change and the change due to the transport of the basis vectors along the curved manifold are expressed. Accounting for the curvature means to include the gravitational field into the description.

From the Christoffels and their first derivatives we then arrive to the Riemann tensor $R^{\mu}_{\alpha\nu\beta}$. The latter is a rank-4 tensor which fully contains all geometric information about the manifold; it is also known as the \textit{curvature tensor}. To the Riemann a rank-2 symmetric tensor is associated known as the Ricci tensor: $R_{\mu\nu}=R^{\alpha}_{\mu\alpha\nu}$. Finally, from Ricci we get a rank-0 tensor (a scalar) $R=R^{\mu}_{\mu}$ denominated the \textit{scalar curvature}. The Ricci tensor and the scalar curvature are used to build the Einstein tensor $G_{\mu\nu}=R_{\mu\nu}-Rg_{\mu\nu}/2$.

The coronation and the center of GR are the famous Einstein equations:
\begin{equation}
G_{\mu\nu}=\frac{\kappa}{2}T_{\mu\nu}
\label{einstein}
\end{equation}
They establish the link between geometry (the Einstein tensor), on the left, and matter/energy as a source of curvature, on the right. The latter is represented by its \textit{energy/momentum} symmetric tensor; the strength of the coupling is expressed by the parameter:
\begin{equation}
\kappa=\frac{16\pi}{c^4}G
\end{equation}

\section{The Schwarzschild solution}
Looking for physical situations where some experimental or observational test of the theory is possible, we are led to consider a simple idealized system consisting of a central static spherical distribution of mass and nothing else. The mentioned spherical symmetry of course refers to space (three dimensions out of four); adding the time independence (static configuration) and going to the full four dimensional representation we actually find a \textit{cylindrical} symmetry (around the time axis). Under these conditions an exact solution for the Einstein equations was found as early as in 1916 by Karl Schwarzschild \cite{schw}.

The Schwarzschild solution in vacuo (i.e. outside the central matter distribution) is expressed by the line element:
\begin{equation}
ds^2=(1-2G\frac{M}{c^2 r})c^2dt^2-(1-2G\frac{M}{c^2 r})^{-1}dr^2-r^2 d\theta^2-r^2\sin^2\theta d\phi^2
\end{equation}
The chosen Schwarzschild coordinates make the spherical rotation symmetry fully visible; $M$ is the total mass of the source and, to shorten a bit the notation, you may introduce the length $\mathcal{M}=G\frac{M}{c^2}$; it is $r\geq R$, being $R$ the radial coordinate delimiting the mass distribution \footnote{We are not interested in black holes here, so we assume also $R>2\mathcal{M}$.}.

\begin{figure}
\begin{center}
\includegraphics[width=.6\textwidth]{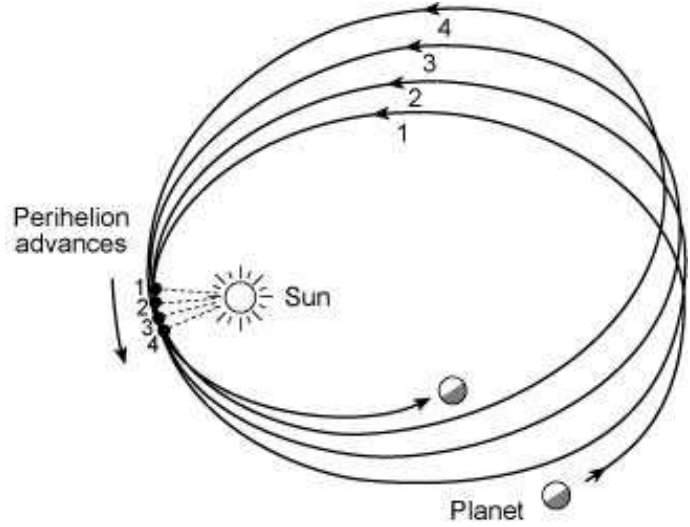}
\caption{Spacely bound orbit in the Schwarzschild space-time surrounding the sun. The precession of the perihelion is clearly visible.}
\label{fig1}
\end{center}
\end{figure}

For our purposes it is interesting to look at the orbits of test particles, i.e. at space trajectories of freely falling objects,  bounded both from below and from above. It is found that no Keplerian orbits exist, but instead non-circular orbits look like the one in fig. \ref{fig1}.
The closest approach point (as well as the farthest) is preceding in the same sense as the rotation of the object: this is the reason why the phenomenon is called 'the perihelion \footnote{or perigee, periastron,... according to what the central body is. The general name is periapsis.} advance'.

It is possible to work out an approximated formula for the angular advance of the perihelion per revolution. If $b$ is the closest approach distance to the center, the angular advance per turn $\Delta \Phi$, to the lowest order in the small parameter $\mathcal{M}/b$, is \cite{perihelion}
\begin{equation}
\delta\phi\simeq6\pi\frac{\mathcal{M}}{b}
\end{equation}
$\delta\phi$ is observable both in the solar system and in binary systems, as we shall see.

\begin{figure}
\begin{center}
\includegraphics[width=1.0\textwidth]{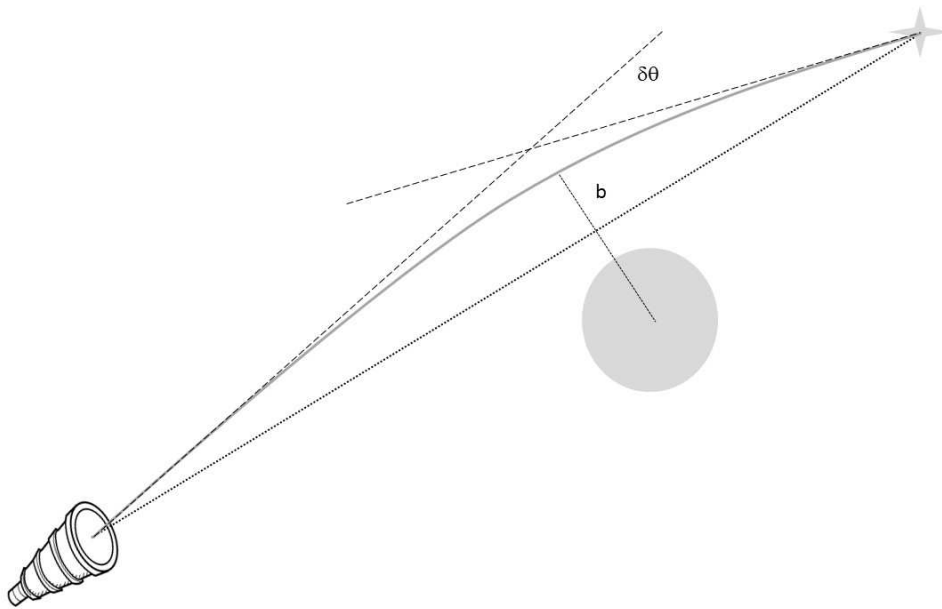}
\caption{Deviation of a light ray in the gravitational field of the sun.}
\label{fig2}
\end{center}
\end{figure}

Another important effect easily worked out in the Schwarzschild geometry is the deviation of light rays due to the presence of the gravitational field. Light 'falls' when approaching a mass and its trajectory is consequently affected like in fig. \ref{fig2}. Such an effect had already been predicted in the framework of Newton's universal gravity and it was possible to evaluate its size in the vicinity of the sun. GR's is however different from Newton's. The expected deviation angle for a light ray passing by a spherical mass is indeed
\begin{equation}
\delta\theta=4\frac{\mathcal{M}}{b}
\label{lente}
\end{equation}
twice as big as the classical value.

There is then one more effect that Einstein predicted even before completing his General Relativity theory \cite{redshift}. In fact, special relativity, photons, mass/energy equivalence and classical energy conservation are enough to see that a light ray climbing up a gravitational potential well must redden. In a limited region close to the surface of the earth (almost uniform gravitational field) the frequency shift (gravitational redshift) $\delta\nu/\nu$ is:
\begin{equation}
\frac{\delta\nu}{\nu}=-\frac{\mathcal{M}}{R_{earth}^2}\,h=-\frac{g}{c^2}\,h
\label{redterra}
\end{equation}
The surface acceleration of gravity on earth $g$ appears in the formula together with the height $h$ of the receiver with respect to the source of light.

In a fully relativistic notation one has that the projection of the four-vector of light $k$ on the four-velocity $u$ of the observer stays constant along the light ray:
\begin{equation}
k\cdot u=g_{\mu\nu}k^{\mu}u^{\nu}=constant
\end{equation}

In practice, for observers holding fixed positions at the emission and absorption points, it is:
\begin{equation}
\frac{\delta\nu}{\nu}=\sqrt{\frac{g_{00(e)}}{g_{00(a)}}-1}
\label{redshift}
\end{equation}

Writing eq. \ref{redshift} in a Schwarzschild space-time, where the observers are respectively at the $R_e$ and $R_a$ radii, one gets:
\begin{equation}
\frac{\delta\nu}{\nu}=\sqrt{\frac{1-2\frac{\mathcal{M}}{R_e}}{1-2\frac{\mathcal{M}}{R_a}}-1}\simeq\frac{\mathcal{M}(R_e-R_a)}{R_e R_a}
\end{equation}
If it is $h=R_a-R_e$ and $h<<R_e\sim R_a\sim R_{earth}$ we come back to (\ref{redterra}).
Assuming that the Schwarzschild solution is a good approximation of the field in proximity of an almost spherical body even when we allow many such bodies to exist in the universe, we may write the expected redshift for a light ray leaving from the surface of a star (whose radius be $R_s$) and arriving on the surface of the earth ($R_\otimes$):
\begin{equation}
\frac{\delta\nu}{\nu}\simeq\frac{\mathcal{M}_s}{R_s}-\frac{\mathcal{M}_\otimes}{R_\otimes}
\end{equation}

\section{Historical tests of GR}
The first historical test met by GR was the anomalous advance of the perihelion of the planet Mercury. While subtracting from the observations all known contributions to the advancement, it remained an unexplained 43'' per Julian century. Einstein's prediction in 1915, according to GR, was precisely 43'' per Julian century.

The next relevant test was with the solar deflection of the light rays. The prediction of GR for light grazing the rim of the sun was, according to formula \ref{lente}, $\delta\theta=1.75"$. This value was verified and validated during a solar eclipse, in 1919, by Arthur Eddington. Even though, with hindsight, we may recognize that the accuracy of Eddington's observation was rather poor, that result was hailed as a triumph for General Relativity, whose fame spread afterwards all over the world.

As for the gravitational redshift, it was initially recognized in the spectral lines of the light emitted by Sirius B \cite{siriusB} in 1925. The first direct measurement on earth was performed in a clever experiment made by Pound and Rebka in 1959. They looked for the red (or blue) shift on the frequency of $\gamma$ rays emitted by $^{57}$Fe when vertically climbing up or down over a distance of 22.5 m in the Jefferson laboratory of the Harvard university. Pound and Rebka exploited the M\"{o}ssbauer effect in order to avoid the recoil of the emitting atom and used a receiver vibrating at acoustic frequencies in order to compensate for the red (blue) shift by means of the corresponding Doppler effect. The expected frequency shift was $\sim2.45\times10^{-15}$ Hz and it was verified with an accuracy of $10\%$; further refinements led the accuracy to $1\%$ and today the gravitational redshift is confirmed to better than 1 part in $10^4$.

\section{Present challenges}
Coming closer to our days we find various areas in which the verification of the predictions of GR are being looked for and, as I have already mentioned, the interest for such verifications has been revived by a number of open problems.

Leaving aside the proper cosmological domain I shall mention, in the following: astrometry and celestial mechanics; gravitational lensing; the equivalence principle; the rotational effects. I will expand a bit on the latter in section \ref{roteff}.

\subsection{Extra-solar celestial mechanics}
An ideal test bed for GR is represented by binary systems containing a pulsar, among which especially the so far unique double pulsar PSR J0737-3029. Such systems are typically composed by one pulsar plus another high density stellar object (white dwarf, neutron star, black hole \footnote{a binary associating a pulsar and a black hole has not been identified until now.}) a few astronomical units apart, or even less than one AU. Gravitational effects are generally much stronger there, than they are in the solar system, so that one expects that monitoring the dynamics of the pair of objects may evidence relativistic phenomena much better than in our planetary system.

An example is given by the periapsis advance, which, in the case of PSR B1913+16 (the famous Hulse and Taylor pulsar) is as big as $4.2 ^\circ$/year, fully compatible with GR\footnote{The 'fully compatible' attribute means that there are various causes concurring to produce the advance and it is not for the moment possible to completely disentangle one from another.} \cite{HT}. In the case of the double pulsar the advance is even $16.9 ^\circ$/year \cite{burgay}.
Other important GR effects visible in binaries are related to the orbital motion and the spin of the compact stars in the pair. I will discuss these effects later on, but it is worthwhile to quote a couple of results here. Both PSR J0737-3029B (a member of the double pulsar) and PSR J1141-6545 \cite{dSP} exhibit de Sitter (or geodetic) precession, i.e. the precession of the spin of the star in the gravitational field of its companion.

A striking correspondence to the predictions of GR is obtained in the case of gravitational waves emission. In fact in vacuo and in the weak field limit, when it is possible to write $g_{\mu\nu}=\eta_{\mu\nu}+h_{\mu\nu}$ with $h_{\mu\nu}<<\eta_{\mu\nu}$, the Einstein equations become approximately \cite[chapter 35]{MTW}:
\begin{equation}
\square \overline{h}_{\mu\nu}\simeq0
\label{wave}
\end{equation}
The square stays for the D'Alembert operator, i.e. the algebraic sum (time has the opposite sign with respect to space) of the four second order partial derivatives $\partial^2/\partial^2 x^{\alpha}$. Eq. \ref{wave} is the typical wave equation for the elements of the tensor:
\begin{equation}
\overline{h}_{\mu\nu}=h_{\mu\nu}-\frac{h_{\alpha}^{\alpha}}{2}\eta_{\mu\nu}
\end{equation}

Eq. \ref{wave} tells us that, in empty space-time, there exist gravitational perturbations propagating like waves at the speed of light \footnote{eq. \ref{wave} is the consequence of a linearization, but there are also some special solutions of the full Einstein equations \ref{einstein} in vacuo propagating like waves.}. The origin of these \textit{gravitational waves} (GW), much as in the case of electromagnetic waves, originates in the accelerated movement of massive bodies\footnote{The whole subject of gravitational waves contains subtleties that we are not discussing here.}. Everybody knows that there are many experiments underway aimed at revealing the presence of gravitational waves; unfortunately however, until now, no positive event has been observed, both because of the expected weakness of the phenomenon and because of the rarity of events producing strong enough pulses.

\begin{figure}
\begin{center}
\includegraphics[width=1.2\textwidth]{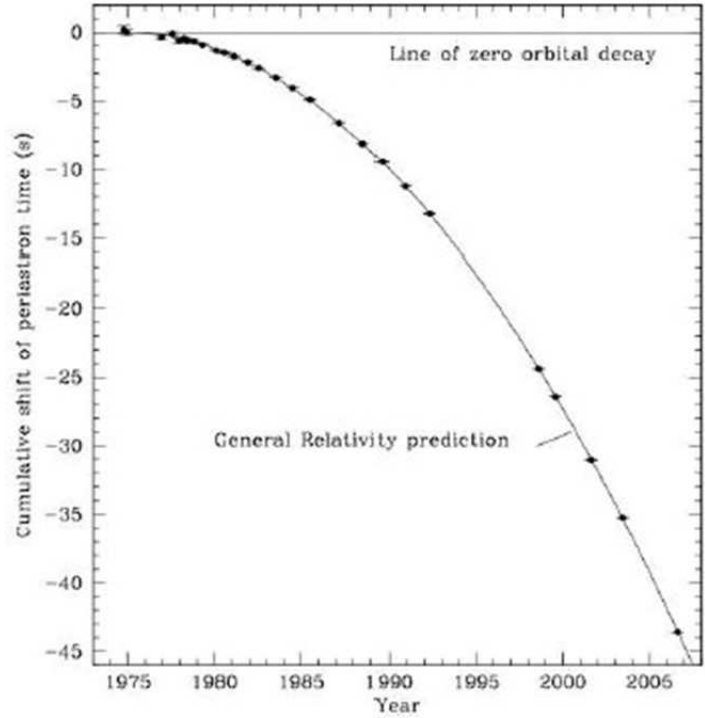}
\caption{Period decay with time in the Hulse and Taylor binary pulsar system.}
\label{fig3}
\end{center}
\end{figure}

On the other hand we know that the emission of GW's should be a rather universal phenomenon. Each gravitationally bound system, made of two or more bodies, is expected to radiate away gravitational energy in the form of waves through its time dependent quadrupole moment. In the same time the potential energy of the system should decrease; in practice, in a pair, the distance between the two bodies should decrease as well as the orbital period. This behaviour has a striking confirmation in various binary pulsars. Fig.\ref{fig3} shows the case of PSR B1913+16 monitored during more than 33 years.

The period decay with time $\dot{P}$, according to the quadrupolar emission of waves, is given by the formula
\begin{equation}
\dot{P}=\frac{192\pi}{5}\Bigl(1+\frac{73}{24}e^2+\frac{37}{96}e^4\Bigr)(1-e^2)^{-7/2}\Bigl(\frac{2\pi \mathfrak{M}}{P}\Bigr)^{5/3}
\label{quadru}
\end{equation}
where $e$ is the eccentricity and
\begin{equation}
\mathfrak{M}=\frac{(m_1m_2)^{2/5}}{(m_1+m_2)^{1/5}}
\end{equation}

The theoretical curve \ref{quadru} fits the observed data of the shift of the periastron time, reported in the graph, within $0.2\%$.

\begin{figure}
\begin{center}
\includegraphics[width=0.5\textwidth]{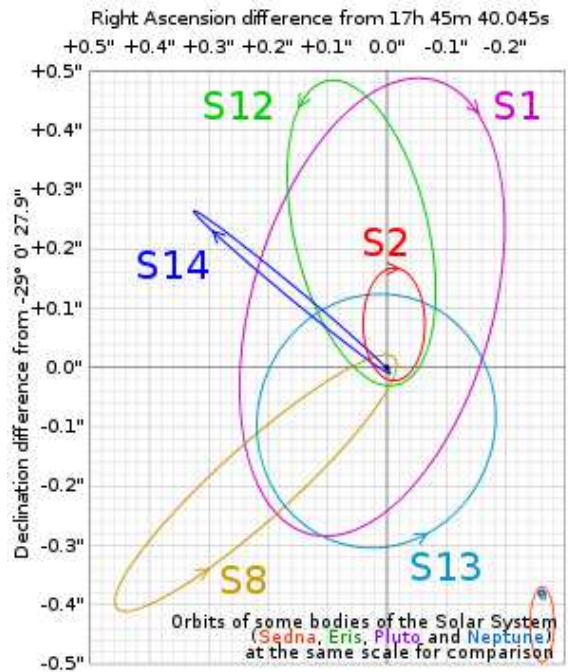}
\caption{Stars surrounding the massive object Sagittarius A*.}
\label{fig4}
\end{center}
\end{figure}

An interesting celestial laboratory for measuring a wealth of relativistic effects is the center of our galaxy, in the Sagittarius constellation. The highest interest comes from a limited region in the middle of which stands an object, named Sagittarius A*, visible only through the gravitational effects it produces on the surrounding stars. The situation is represented in fig. \ref{fig4}.

The orbits of five, out of more than 20 stars, observed during over 15 years by the Keck infrared telescope, are shown. The only actually complete orbit so far is the one of S2; the others are partly observed and partly reconstructed. Sagittarius A* sits in the origin of the figure; its estimated mass is $\sim4.3$ million solar masses \cite{mass}, its volume has only upper bounds, but the object is currently deemed to be a supermassive black hole as the ones which are expected to lie in the center of most galaxies.

In the case of Sagittarius A*, as well as for the binary pulsars, while time goes on and the observation continues, data cumulate and the statistics become better and better, so that the accuracy with which all relativistic effects influencing the involved mechanics are measured will grow year after year.

\subsection{Lensing and astrometry}
As I have already recalled above, the deviation of light rays induced by gravity has been the phenomenon that marked the passage of GR from a theory for a few experts to a mass phenomenon. Today one speaks of \textit{gravitational lensing} which is observed everywhere in the sky where large masses, such as galaxies or galaxy clusters, are interposed on the way of light from faraway background sources.

Regarding lensing it is worth mentioning an underway space mission, named GAIA. It consists of a spacecraft carrying a pair of telescopes and was launched in December 2013. GAIA is now sitting in the Lagrangian L2 point between the earth and Mars, and is gathering data. The main purpose of the mission is to build a new stellar catalog observing approximately one billion stars of our galaxy, but the important point is that the accuracy in the celestial coordinates of each object will be some units of $\mu$arcsec, i.e $\sim10^{-11}$rad. Such an accuracy makes the deviations of the light rays by the gravitational fields in the solar system important, not only in the case of the sun (whose influence at that scale extends well beyond the inner solar system) but also from other bodies such as Jupiter, Saturn or even Mars. In practice the elaboration and use of the data collected by GAIA, requires an accurate modelling of the relativistic effects present in the solar system, thus resulting indirectly in an accurate test of the theory.

\subsection{The equivalence principle}
The equivalence principle in relativity has various formulations. The one which has been used most for direct verification is antecedent to relativity, going back to Galileo, and maintains that the inertial and the gravitational mass of any object coincide exactly. Of course experimentally one may only set upper limits to the equality between the two masses and the actual limits have been progressively improved while adopting more and more refined and accurate techniques. To day measurements done on earth by means of torsion balances tell us that the inertial and the gravitational mass do coincide up to a few parts in $10^{14}$. Tighter constraints can be pursued by a number of experiments in space that have been proposed, like STEP (Satellite Test of the Equivalence Principle) \cite{STEP}, and GG (Galileo Galilei) \cite{GG}), aiming respectively to an accuracy of $10^{-18}$ and $10^{-17}$. Under development is MICROSCOPE (MICROSatellite \`{a} train\'{e}e Compens\'{e}e pour l'Observation du Principe d'Equivalence) \cite{micro}, whose objective is also an accuracy of $10^{-18}$; the mission could fly in 2015-16.

\section{Gravitomagnetism}
\label{GM}
A whole class of phenomena predicted by Einstein's theory has to do with moving (especially rotating) sources of gravity. Let us write the general form of the line element of the space-time surrounding a massive body:
\begin{equation}
ds^2=g_{00}c^2dt^2+g_{ij}dx^i dx^j+2g_{0i}cdtdx^i
\label{lineabase}
\end{equation}
The equivalence principle tells us that it is always possible to choose the coordinates so that the $g_{\mu\nu}$'s are converted into the Minkowski $\eta_{\mu\nu}$'s at a given space-time position.  It is however in general not possible to find a \textit{global} coordinate transformation converting $g_{\mu\nu}$ into $\eta_{\mu\nu}$ everywhere, but it is also assumed that space-time be asymptotically flat, i.e. the metric tensor tends to the Minkowski form far away from the source.

The elements of the metric tensor are functions of the coordinates and the mixed terms $g_{0i}$ imply that the source of gravity is moving with respect to the observer (most often the latter is assumed to be at rest in the chosen reference frame). By the way, considering any path in space-time, the contribution $g_{0i}cdtdx^i$ to the line element has a sign which depends on the orientation of the space projection of the world-line with respect to the axes of the reference frame.

It is known that the world-line of a freely falling test particle is described by a geodetic line. The latter may be written so that the second order derivatives of the coordinates with respect to $s$, which are proportional to the proper acceleration of the object, are singled out; in the following the space components are reproduced:
\begin{equation}
\frac{d^2 x^i}{ds^2}=-\Gamma^i_{00}(u^0)^2-\Gamma^i_{jk}u^j u^k-2\Gamma^i_{0j}u^0u^j
\label{geod}
\end{equation}
It is $u^{\alpha}=dx^{\alpha}/ds$.

Eq. \ref{geod} may be read also as the 'gravitational force' per unit mass divided by $c^2$, or the gravitational acceleration ($/c^2)$ in the given reference frame. The first term on the right hand side does not depend on the space components of the four velocity (which are in turn proportional to the components of the ordinary velocity) and in practice corresponds, in the weak field limit, to the Newtonian gravity acceleration divided by $c^2$. The next term, proportional to second degree products of the space components of the four-velocity, can be interpreted as a sort of 'viscous' term \cite{viscous} in analogy with the fluidodynamics viscous forces (even though here it is not garanteed that the effect be against motion).

The last term of eq. \ref{geod} deserves a bit more attention. Writing it \textit{in extenso} we get:
\begin{equation}
-2\Gamma^i_{0j}u^0u^j=-g^{i\alpha}\Biggr(\frac{\partial g_{0\alpha}}{\partial x^j}+\frac{\partial g_{\alpha j}}{\partial x^0}-\frac{\partial g_{0j}}{\partial x^{\alpha}}\Bigg)
\end{equation}
Let us introduce the simplifying assumption that the metric elements do not depend on time (stationary gravitational field), so that the formula becomes a bit simpler:
\begin{equation}
-2\Gamma^i_{0j}u^0u^j=-g^{i\alpha}\Biggr(\frac{\partial g_{0\alpha}}{\partial x^j}-\frac{\partial g_{0j}}{\partial x^{\alpha}}\Bigg)
\label{rotore}
\end{equation}

Identifying now the three $g_{0i}$ components of the metric tensor with the three components of an ordinary $H_i$ vector we easily recognize in the brackets of eq. \ref{rotore} a component of the ordinary three-dimensional curl of $\overrightarrow{H}$. In analogy with electromagnetism we shall call $\overrightarrow{H}$ \textit{gravitomagnetic vector potential} and $\overrightarrow{B}_g=c\overrightarrow{\nabla}\wedge\overrightarrow{H}/2$ will be the \textit{gravitomagnetic field}.

As the next step let us consider a (quite common) situation in which the velocity of the falling object is much smaller than the speed of light, $v^i<<c$, so that $u^0\simeq 1$, $u^i\simeq v^i/c$ and second order terms, as the ones in the 'viscous' term mentioned above, may freely be neglected. Multiplying by the (rest) mass $m$ of the test particle, looking for the covariant components and making the summations explicit, we obtain from the third term of eq. \ref{geod} the force:
\begin{equation}
F_i=2m(v_jB_k-v_kB_j) \rightarrow \overrightarrow{F}=2m\overrightarrow{v}\wedge\overrightarrow{B}_g
\label{lorentz}
\end{equation}

In practice we see that a moving mass in a reference frame where $g_{0i}\neq 0$ feels an additional gravitational force \ref{lorentz} much like the Lorentz magnetic force; the coupling parameter in the electromagnetic case is the charge $q$, in gravitomagnetism it is twice the rest mass $2m$.

The last remark of this section is that sometimes the presence of non-null off-diagonal time-space terms in the metric tensor can be an effect of the choice of the coordinate system, so that a change of the coordinates suffices to wash them away. However if the source of gravity is rotating it is not possible, by any coordinate change to have simultaneously a null $\overrightarrow{H}$ and a static (i.e. time independent) metric. This fact tells us that the typical environment where to look for gravitomagnetic effects is the space-time surrounding a rotating mass.

\section{Rotation effects}
\label{roteff}
The typical line element of the space-time containing a spherical mass in steady rotation may usefully be written using spherical coordinates in space:
\begin{equation}
ds^2=g_{00}c^2dt^2+g_{rr}dr^2+g_{\theta\theta}d\theta ^2+g_{\phi\phi}d\phi ^2+2g_{0\phi}cdtd\phi
\end{equation}
Considering the symmetry, the $g_{\mu\nu}$'s depend neither on $t$ nor on the azimuthal angle $\phi$.

Within the solar system (and almost everywhere in the universe) the weak field conditions hold, so that the line element can be approximated to first order in the small parameters becoming:
\begin{equation}
ds^2\simeq\Biggl(1-2\frac{\mathcal{M}}{r}\Biggr)c^2dt^2-\Biggl(1+2\frac{\mathcal{M}}{r}\Biggr)dr^2-r^2 d\theta ^2-r^2 sin^2\theta d\phi ^2+4\frac{j}{r^2}r sin\theta cdt d\phi
\label{weak}
\end{equation}
The last term has been written evidencing separately the dimensionless part of the mixed element of the metric tensor and the components of the line element. The parameter $j$ has the dimension of a squared length and is related to the angular momentum of the source $J$ through:
\begin{equation}
j=\frac{G}{c^3}J
\end{equation}

The components of the gravitomagnetic vector potential are easily read out of \ref{weak}. Actually only one component differs from zero:
\begin{equation}
H_{\phi}=2\frac{j}{r^2}sin\theta
\label{dipole}
\end{equation}
In full three-dimensional vector notation and being $\hat{r}$ the unit vector along the radial direction, it is:
\begin{equation}
\overrightarrow{H}=-2\frac{\overrightarrow{j}\wedge\hat{r}}{r^2}
\end{equation}

Looking at \ref{dipole} it is easy to recognize a dipole potential. From it one obtains a dipolar gravitomagnetic field:
\begin{equation}
\overrightarrow{B}_g=-2\frac{G}{c^2r^3}\Bigl[\overrightarrow{J}-3\bigl(\overrightarrow{J}\cdot\hat{r}\bigr)\hat{r}\Bigr]
\end{equation}

In practice we see that $\overrightarrow{B}_g$ for a steadily rotating mass looks like the external magnetic field of a bar magnet.

The presence of $\overrightarrow{B}_g$ leads to some interesting consequences. First of all the magnetic analogy tells us that a test mass endowed with intrinsic angular momentum $\overrightarrow{S}$ (in practice a test gyroscope) will let its own rotation axis precede about the local direction of the gravitomagnetic field because of the torque exerted on it by gravity. In fact the torque (derivative of $\overrightarrow{S}$ with respect to the proper time $\tau$), the angular momentum $\overrightarrow{S}$ and the precession angular speed $\overrightarrow{\Omega}$ are related to each other by the classical equation of motion of the gyroscope:
\begin{equation}
\frac{d\overrightarrow{S}}{d\tau}=\overrightarrow{\Omega}\wedge\overrightarrow{S}
\end{equation}

Explicitly solving the equation for a gravitational field yields  \cite[chapter 40]{MTW}:
\begin{equation}
\overrightarrow{\Omega}=-\frac{1}{2c^2}\overrightarrow{v}\wedge\overrightarrow{a}+\frac{3}{2}\overrightarrow{v}\wedge\overrightarrow{\nabla}\Phi
-\frac{c}{2}\overrightarrow{\nabla}\wedge\overrightarrow{H}
\label{gyr}
\end{equation}

The first term on the right is a special relativity effect known as the Thomas precession: it depends on the velocity $\overrightarrow{v}$ of the gyro and its non-gravitational acceleration $\overrightarrow{a}$. The second term manifests the coupling of the velocity with the Newtonian gravitational force ($\Phi$ is the gravitational potential), which we may call the \textit{gravitoelectric} force: this term is also known as the geodetic or de Sitter precession. Finally the third term coincides with the local gravitomagnetic field $\overrightarrow{B}_g$ and is responsible for the so called Lense-Thirring drag, manifested here in the form of an additional precession of the gyroscope.

There is another important effect related to the presence of a gravitomagnetic field. It has to do with the propagation of light\footnote{Actually the same effect appears also on any other signal or object moving along a closed space trajectory, provided the local propagation speed is the same at any position along the path \cite{any}} along a closed space path. In the case of light, the line element \ref{lineabase} is identically zero, so it is possible to solve it for the coordinated time interval $dt$, obtaining:
\begin{equation}
dt=\frac{1}{c g_{00}}\Biggl(-g_{0i}dx^i+\sqrt{g_{0i}^2(dx^i)^2-g_{00}g_{ij}dx^idx^j}\Biggr)
\label{tempo}
\end{equation}
The sign in front of the square root has been chosen so that, while moving along the trajectory either to the right or to the left, times flows always towards the future.

If now we wish to evaluate the time of flight of a light ray along a closed contour, we may integrate \ref{tempo} along the path. Since the first term, linearly depending on the $dx^{\,i}$'s, has different signs according to the propagation sense, two different final results are obtained for the two opposite senses. The difference between the two travel times, expressed in term of \textit{proper} time of an observer sitting at the start/arrival point, will be:
\begin{equation}
\delta\tau=\tau_+-\tau_-=-\frac{2}{c}\sqrt{g_{00}}\oint\frac{g_{0i}}{g_{00}}dx^i
\label{delta}
\end{equation}

The asymmetric propagation is due to the presence of non-zero off-diagonal time-space elements in the metric tensor, whatever the origin of their presence be. In flat space-time $\delta\tau$ is due to the motion of the source/receiver of light along a closed path: this is the \textit{kinematical} Sagnac effect \cite{sagnac}. In a curved space-time originated by a rotating mass both the de Sitter coupling to the gravitoelectric field and the gravitomagnetic field contribute additional \textit{physical} terms to the time of flight difference.

\section{Experiments on gravitomagnetism}
The relativistic precession of a gyroscope in the gravitomagnetic field of a rotating mass has qualitatively been observed in binary pulsars (especially the double pulsar) where the 'gyroscope' is both either of the two spinning companions or the orbit of one of the two stars around the center of mass of the pair. The results, so far, are in general compatible with GR. It is not yet possible to say more than that because, as I have already written, the number of parameters entering the dynamics of a binary system is high and it is not easy to separate the effect of each of them. The accuracy is however growing in time while observation goes on, so: wait and see.

Still in the domain of observation lies the behavior of a body closer to us: the moon. The whole orbit of the moon around the earth may be thought of as a gigantic gyroscope dragged to precede by the gravitomagnetic field of the earth. The orbit of the moon is currently monitored by laser ranging from the earth. There has been a claim that the data from the lunar laser ranging confirm the Lense-Thirring drag with the extremely good accuracy of $0.1\%$ \cite{MNT}, but the result is controversial \cite{Kope}.

\subsection{Gravity Probe B}
\begin{figure}
\begin{center}
\includegraphics[width=0.8\textwidth]{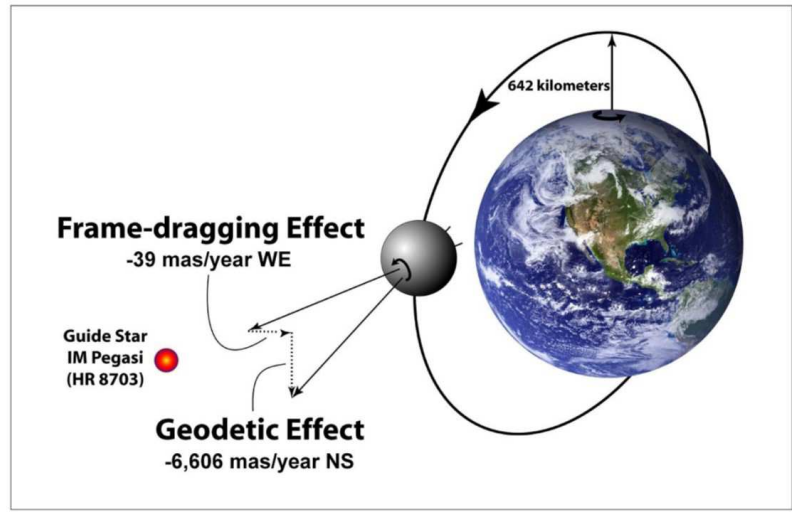}
\caption{The GP-B mission and its findings.}
\label{fig5}
\end{center}
\end{figure}

Coming to direct experiments in space, an important endeavor has been Gravity Probe B (GP-B), launched and performed by NASA and the Stanford University in 2004 and taking data until September 2005. The quantity to measure was the precession angle of the axis of four gyroscopes carried by a spacecraft in polar orbit around the earth (see fig.\ref{fig5}). Being the satellite in free fall the Thomas precession term in \ref{gyr} is zero; the only active contributions remain the general relativistic ones. The experiment has verified the geodetic precession within $0.28\%$, but unfortunately uncontrolled spurious torques concentrated on one of the gyros have spoiled part of the results and the Lense-Thirring frame drag could only be tested at a disappointing $19\%$ accuracy \cite{GPB}.

\subsection{The LARES mission and its antecedents}
\begin{figure}
\begin{center}
\includegraphics[width=0.8\textwidth]{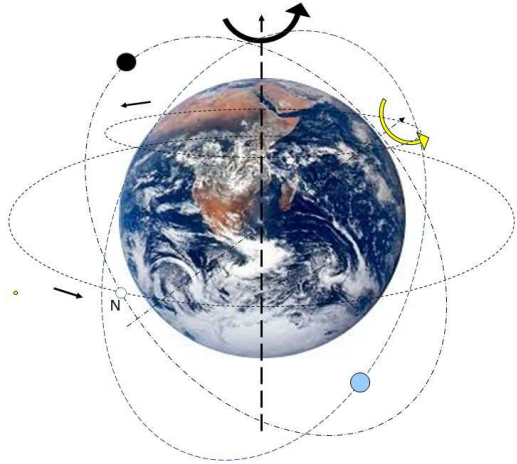}
\caption{The orbits of the two LAGEOS satellites are represented. The axis of their orbits precedes around the rotation axis of the earth and, by consequence, the nodes $N$ advance along the equator at each revolution. For simplicity only one node and one orbital axis are shown.}
\label{fig6}
\end{center}
\end{figure}

The frame dragging by the earth has also been measured before the completion of GP-B exploiting the laser ranging of the LAGEOS satellites.
The two LAGEOS were launched respectively in 1976 and in 1992 for geodesy purpose. The possibility to use their orbits in order to verify the Lense-Thirring effect was initially proposed by Ciufolini \cite{ciufo1}. An orbiting satellite can indeed be considered as a gyroscope and its axis (perpendicular to the plane of the orbit) will be dragged to precede around the axis of the earth, i.e. the dipolar source of the gravitomagnetic field. The observable phenomenon will be an advancement of the nodes of the orbit of the satellite. The idea is represented in fig.\ref{fig6}.

The delicate point with this type of experiment is that you need to know with the highest accuracy the gravitoelecric field of the earth in order to eliminate the causes of advancement of the nodes depending on the non-sphericity and non-homogeneity of our planet. The required knowledge is acquired and improved by means of independent and dedicated missions such as NASA's GRACE (GRAvity recovery and Climate Experiment) and ESA's GOCE (Gravity and Ocean Circulation Experiment); the better their results are, the better becomes the determination of the gravito-magnetic advancement of the nodes of the orbits of satellites.

Analyzing the available data of the laser ranging of the orbits it was initially possible to verify the general relativistic drag with an accuracy in the order of $30\%$. Afterwards, using a better modelling of the gravito-electric field of the earth based on the data of the GRACE mission, the accuracy was improved to $10\%$ \cite{ciufo2011}.

\begin{figure}
\begin{center}
\includegraphics[width=0.4\textwidth]{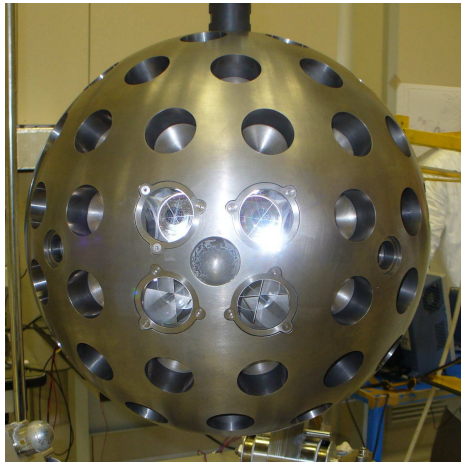}
\caption{A picture of the LARES satellite before being launched.}
\label{fig7}
\end{center}
\end{figure}

The two LAGEOS were not designed and launched to test GR, but now a dedicated mission, based on the same principle, is underway. The LARES (LAser RElativity Satellite) satellite was launched on February 2012 and its orbit is being monitored by laser ranging in order to determine the advancement of its nodes \cite{ciufo3}. LARES is following a trajectory which is closer to a geodetic than in the case of the LAGEOS. This happens because LARES is closer to a test particle. It is a monolithic sphere, $36.4$ cm in diameter, made of a tungsten alloy, carrying on its surface 96 retroflectors for the laser light sent from the ground (see fig.\ref{fig7}). The measurement of the Lense-Thirring drag is expected to be carried within a few $\%$ accuracy.

\section{Ring lasers}
\begin{figure}
\begin{center}
\includegraphics[width=0.8\textwidth]{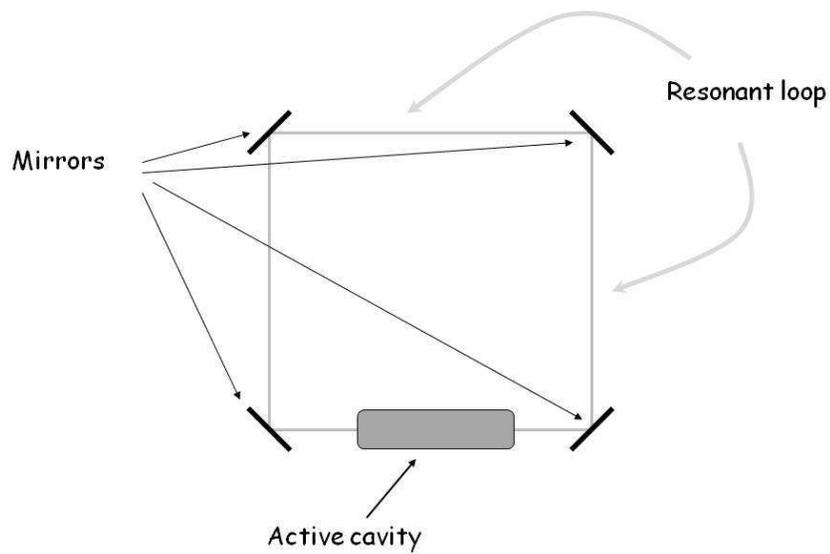}
\caption{Principle scheme of a square ring-laser.}
\label{fig8}
\end{center}
\end{figure}

An interesting device to be considered when looking for possible asymmetries in the propagation of light is a ringlaser. The scheme of the apparatus is shown in fig.\ref{fig8}.

 Coming to the second phenomenon, mentioned at the end of section \ref{roteff}, we have seen that in presence of generalized rotations there is an asymmetric propagation of light along closed paths. The difference in the times of flight can lead to interference phenomena. In a ringlaser however we have a continuous emission of light of a given fiducial frequency (the proper frequency of the lasing cavity) and in practice in the physical ring two stationary light beams exist, one travelling to the right and the other to the left. Each beam has to be coherent with itself, which means that it resonates over a wavelength (then a frequency) which covers the ring an integer number $N$ of times. In terms of times of flight, however, the difference between clockwise and counter-clockwise produces two slightly different resonation frequencies (or wavelengths) referred to the same integer $N$ \footnote{This is true in the \textit{fundamental mode} and we shall assume that the laser works in that mode.}.

 The difference in frequencies is proportional to the time of flight difference. According to what we have stated above it must indeed be $t_+=NT_+$ and $t_-=NT_-$, where $T_+$ and $T_-$ are respectively the periods of the counterclok- and clock-wise waves. They are forced to be different in order to account for the same $N$ with different times of flight. Once we remember that the frequency is the inverse of the period, the following sequence of steps leads to the final proportionality relation:
 \begin{equation}
 \delta t=N(T_+-T_-)=N(\frac{1}{\nu_+}-\frac{1}{\nu_-})=N\frac{\nu_--\nu_+}{\nu_+\nu_-}
 \end{equation}
 Approximating the product $\nu_+\nu_-$ with the square of the proper frequency of the laser $\nu^2$ and introducing the wavelength $\lambda$ and the length of the ring $P$ the final formula is:
 \begin{equation}
 \delta t\simeq N\frac{\delta\nu}{\nu^2}=N\lambda\frac{\delta\nu}{c^2}\lambda=P\frac{\delta\nu}{c^2}\lambda
 \label{batti}
 \end{equation}

 Now, when two oppositely propagating waves with slightly different frequencies superpose, a beat is produced. The beat frequency equates half the difference between the two mentioned frequencies. In practice the power density of the compound wave is modulated at the beat frequency. Spilling a bit of the wave, for instance at a corner of the square in fig. \ref{fig8}, it is possible to read out the beat frequency, which, as we have seen, is proportional to the times of flight difference. Thinking of an experiment on earth we must still consider that we need to use the proper lab time, but also that the coordinates in our lab are not the ones used, say, in formula \ref{weak}. From those coordinates we have to go to a reference frame co-rotating with the earth and our fiducial observer shall coincide with an inertial observer instantly co-moving with the laboratory (in practice we have to perform a Lorentz transformation in addition to the rotation of the axes). To the lowest approximation order in the small relevant quantities the elements of the metric tensor of our interest, in geographic coordinates, become:
 \begin{eqnarray}
 g_{0\phi}&\simeq& \Bigl[2\frac{j}{r}-\Bigl(r^2+2\mathcal{M} r\Bigr)\frac{\Omega}{c}\Bigr]\sin^2\theta \nonumber \\
 g_{00}&\simeq& 1-2\frac{\mathcal{M}}{r}-\frac{\Omega^2r^2}{c^2}\sin^2\theta \nonumber
 \end{eqnarray}

 The numerical values of the parameters, close to the surface of the earth, are:
 \begin{eqnarray}
 \mathcal{M}&=&G\frac{M}{c^2}\simeq 4.4\times 10^{-3} \,\textrm{m} \nonumber \\
 j&=&G\frac{J}{c^3}\simeq 1.75\times 10^{-2} \,\textrm{m}^2 \nonumber \\
 \Omega&\simeq& 7.27\times 10^{-5} \,\textrm{rad/s} \nonumber
 \end{eqnarray}

 Summing up, the expected signal to be detected by a ringlaser will be \cite{gingerpr}:
 \begin{equation}
 \delta\nu=4\frac{A}{\lambda P}\Bigl[\overrightarrow{\Omega}-2\frac{\mathcal{M}}{R}\Omega\sin\theta\,\hat{u}_{\theta}+c\frac{j}{R^3}(2\cos\theta\,\hat{u}_r+\sin\theta\,\hat{u}_\theta\Bigr]\cdot\hat{u}_n
\label{signal}
 \end{equation}
$R$ is the (average) radius of the earth.
The terms multiplying the square bracket form the \textit{scale factor}; the bigger it is the stronger is the signal. Increasing the size of the ring apparently goes in the direction of increasing the sensitivity, provided mechanical instabilities do not set in spoiling the advantage of the size. The shape of the ring is not important, excepting the requirement of a convenient area/perimeter ratio; the best such ratio is obtained by a circle, however practically this implies the use of an optical fiber, which means propagation in a dense material that cannot be perfectly homogeneous and stable.

The first term in the square brackets of eq.\ref{signal} accounts for the kinematical Sagnac effect. The second term represents the geodetic (or de Sitter) precession rate. Finally the third term contains the gravitomagnetic effect. The two physical terms, on earth, are $\sim10^{-9}$ times the Sagnac term.

\subsection{Existing devices}
According to the description given in the previous section, ringlasers are interesting rotation sensors and actually the practical use that has been made of them, since the '70s of the last century, is precisely that. A commercial ringlaser is a device most often not bigger than a hand, including or not an optical fiber, used as rotation sensor on board of planes, ships, submarines, missiles... They replace the old mechanical gyroscopes: this is why they are often called \textit{gyrolasers}. They have no moving parts, an extremely low energy requirement and a high sensitivity. The latter can arrive to $\sim 10^{-7}$rad/\,s/$\surd Hz$, on the edge of the diurnal angular velocity of our planet.

\begin{figure}
\begin{center}
\includegraphics[width=0.8\textwidth]{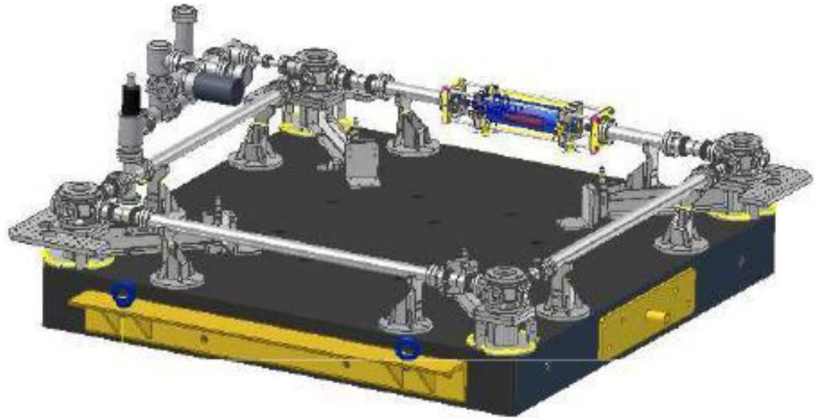}
\caption{The figure presents the G-Pisa square ring.}
\label{fig9}
\end{center}
\end{figure}

Another type of higher sensitivity ringlasers is destined to research purposes, especially in geophysics. Usually these devices are bigger and do not use optical fibers. The whole loop acts as a resonating cavity and light travels in vacuo, or, to say better, in low pressure inert gas. If a He-Ne laser is used the pipe containing the beams is filled with He vapor. An example is the G-Pisa ring, schematically represented in fig.\ref{fig9},
G-Pisa was a square ring initially designed for monitoring the tiny rotations affecting the Virgo interferometer central area. The side of the square loop was $1.35$ m long; a He-Ne laser was used. In order to insure an adequate rigidity and stability of the ring, its four mirrors were mounted on a monolithic granite table, that could be tilted in order to evidence rotations around different axes \cite{gpisa}. After its service at Virgo, G-Pisa has been moved at the Pisa laboratories of INFN and used to test various technical aspects of ringlasers. The finally attained sensitivity was in between $\sim10^{-9}$ and $10^{-10}$ rad/ s/$\surd Hz$.

Very large research ringlasers in various configurations were built in New Zealand, in the underground laboratory of the Cashmere Cavern, near Christchurch. The reason for the underground location is that the sensitivity was so high to allow the devices to sense any kind of rotational noise induced by surface solicitations, such as wind, pressure changes, mass displacement on the ground... Some of the rings arrived to 20 m long side so considerably enhancing the scale factor. Unfortunately, the inability to give a rigid configuration to such big structures negatively compensated the increased scale factor, by injecting uncontrolled mechanical instability in the form of rotational noise. Today the laboratory of the University of Canterbury at Christchurch in the Cashmere cavern remains unusable after an earthquake that seriously damaged it in February 2011.

\begin{figure}
\begin{center}
\includegraphics[width=0.8\textwidth]{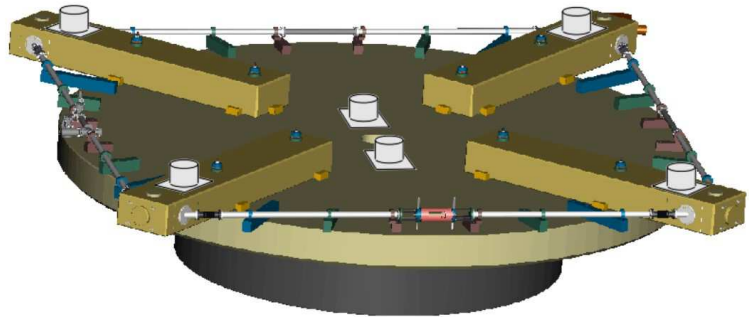}
\caption{The G ring in Wettzell.}
\label{fig10}
\end{center}
\end{figure}

The presently best ringlaser in the world is the Gro{\ss} Ring G at the German Geod\"{a}tisches Laboratorium in Wettzell (Bavaria). It is a square ring, 4 m in side, mounted on a monolithic table made of zerodur, an extremely rigid and thermally stable ceramic material. A schematic view of G is presented in fig.\ref{fig10}. The table rests on a pillar rooted in the solid rock bed beneath the laboratory. The instrument works under a pressure vessel in constant temperature regime; the whole laboratory is located under an artificial mound, 30 m thick, providing a reasonable isolation from surface disturbances.

The performance of G is excellent, reaching now a few $10^{12}$ rad/ s/$\surd Hz$ \cite{wet}, which is close to the capacity to detect the GR terms of \ref{signal}. Actually G, from underground, feels the small diurnal wobbling of the orientation of the terrestrial axis, not to speak of the terrestrial tides in the earth crust, the global movement of the surface of the oceans, etc.

\section{GINGER}
\begin{figure}
\begin{center}
\includegraphics [width=0.8\textwidth]{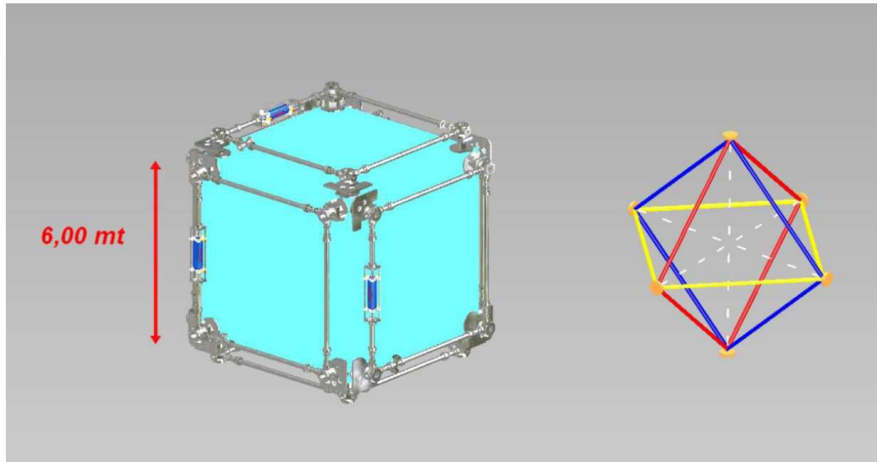}
\caption{Two possible alternative configurations of the three-dimensional array of square rins of GINGER.}
\label{fig11}
\end{center}
\end{figure}

The extremely interesting results of G have inspired the GINGER (Gyroscopes IN GEneral Relativity) project. GINGER will be a three-dimensional array of square laser rings. Since the purpose is to measure the physical terms of the precession rate in \ref{signal} (and in particular the gravitomagnetic Lense-Thirring effect), being all effective rotations vectors, the detection of all three components thereof is required, hence the thee-dimensions of the device. In order to have a good scale factor, the length of the sides of the rings will be not less than 6 m. Fig.\ref{fig11} shows two possible configurations of GINGER.
The cube would provide redundancy, having six square loops; the more elegant octahedron would have only three mutually orthogonal squares, but would offer a better control of the geometry. The 'monument', i.e. the supporting frame, of GINGER will be made of concrete, following a strategy different from the one of G-Wettzell: instead of relying on structural rigidity, the stability and control of the geometry will be pursued dynamically. Fabry-P\'{e}rot cavities will be placed along the diagonals of all squares giving a real time feedback signal controlling the position of the corner mirrors, by means of piezoelectric actuators.
The design quality factor of the cavities will be better than $3\times 10^{12}$ and the target accuracy for the Lense-Thirring effect will be $1\%$ or better with a one year integration time. GINGER will be underground and its final location will be the Gran Sasso Laboratories.

At the present stage of development of GINGER two intermediate devices exist: GP-2 and GINGERino. They are both square rings and are offsprings and heirs of G-Pisa. GP2 ($1.6$m side) is located in Pisa and is used to test the dynamical control of the geometry technique \cite{gp2}. GINGERino (4m side) is located in the Gran Sasso Laboratories and is monitoring the site, from the viewpoint of rotational noise, temperature and pressure stability etc.

More technical details on the GINGER project are given in another talk of the present proceedings.

The GINGER collaboration has as principal investigator Angela Di Virgilio of the PISA INFN section and includes: in Italy, the Pisa, Padua and Legnaro INFN groups, the CNR SPIN insitute in Naples, the universities of Padua, Pisa, Naples, Siena and the Politecnico di Torino; in Germany, the Technische Universit\"{a}t M\"{u}nchen; and in New Zealand, the University of Canterbury in Christchurch.

\section{Matter waves}
It is worth mentioning, as I already did in a footnote of section \ref{roteff}, that the Sagnac effect, even in its general form including GR effects, is not limited to light waves, but happens also with other types of waves. A possibility indeed exists to use matter waves beams. In the case of matter the relevant undulatory quantity is deBroglie's wavelength associated with the momentum of a particle:
\begin{equation}
\lambda=\frac{mv}{\hbar}
\end{equation}

$\lambda$ for a particle can be much smaller than the wavelength of visible light: from {\AA}'s to thousands of {\AA}'s. Considering the scale factor in \ref{signal} we see that we may easily gain a factor $1000$ using particle beams rather than light. You may think to split a matter beam in two parts by means of a diffracting crystal, then, always using diffracting gratings, let the two half beams follow different paths (of equal length) and finally superpose again. If the whole apparatus rotates (and/or a gravitomagnetic field is present) an asymmetric propagation will happen and will show up in the form of a phase difference $\Delta\Phi$ detectable by interferometric techniques. The expected phase difference, written evidencing the frequency (i.e. the mass) of the particles rather than the wavelength, depends on the angular velocity $\overrightarrow{\Omega}$ according to:
\begin{equation}
\Delta\Phi=4\pi\frac{m}{h}A\overrightarrow{\Omega}\cdot\hat{u}_n
\end{equation}

Unfortunately the advantage of much shorter wavelengths is at present compensated by the need to have much smaller devices than in the case of light. Furthermore the stability and the coherence of matter waves are much worse than with laser beams. For the moment, particle beams are not competitive with ringlasers, but, of course, technologies go on improving.

For completeness I must also mention the possibility to use macroscopic quantum systems, such as superconductors or superfluids. In superconductors electromagnetic phenomena are of course dominant, but superfluids (such as $^3He$ or $^4He$) have no such problems. The idea of using a superfluid current in a ring, more or less as in a SQUID, has been considered and some experiments have also already been made \cite{sato}. Again, as for now, the technique is interesting but not competing with ringlasers.

\section{Conclusion}

I have reviewed some relevant examples of actual or under implementation experiments in fundamental gravity. As we have seen, the domain of experimental General Relativity is vivacious and promising. Besides the possibilities offered by precision astrophysics and astrometry, important experimental opportunities are given also on the earth, or in the surrounding space. Special attention deserves the use of light as a privileged and intrinsically relativistic probe for gravity and the structure of space-time. We have in particular seen the importance of ringlasers and we have devoted a specific attention to the GINGER experiment, now under development for the Gran Sasso Laboratories. The expected improvements in the laser technologies, as well as in the matter beams management and in superfluid based devices, promise a near future dense of exciting results.

\end{document}